\documentclass[twocolumn,tighten,times]{aastex61}

\newcommand{\psr}{J1846$-$0258}
\catcode`\@=11
\newcommand{\gapprox}{\mathrel{\mathpalette\@versim>}}
\newcommand{\lapprox}{\mathrel{\mathpalette\@versim<}}
\newcommand{\propapprox}{\mathrel{\mathpalette\@versim\propto}}
\newcommand{\@versim}[2]
  {\lower3.1truept\vbox{\baselineskip0pt\lineskip0.5truept
\ialign{$\m@th#1\hfil##\hfil$\crcr#2\crcr\sim\crcr}}}
\catcode`\@=12

\defcitealias{rc84}{RC84}

\shorttitle{EXPANSION AND FADING IN KES 75 PWN}

\begin{document}

\title{Expansion and Brightness Changes in the Pulsar-Wind Nebula in
  the Composite Supernova Remnant Kes 75}

\email{reynolds@ncsu.edu}

\author{Stephen P. Reynolds}
\affiliation{Department of Physics, North Carolina State University, 
Raleigh, NC 27695-8202, USA}

\author{Kazimierz J. Borkowski}
\affiliation{Department of Physics, North Carolina State University, 
Raleigh, NC 27695-8202, USA}

\author{Peter H. Gwynne}
\affiliation{Department of Physics, North Carolina State University, 
Raleigh, NC 27695-8202, USA}

\begin{abstract}

We report new {\sl Chandra} X-ray observations of the shell supernova
remnant (SNR) Kes 75 (G29.7$-$0.3) containing a pulsar and pulsar-wind
nebula (PWN). Expansion of the PWN is apparent across the four epochs,
2000, 2006, 2009, and 2016.  We find an expansion rate between 2000
and 2016 of the NW edge of the PWN of $0.249\% \pm 0.023\%$ yr$^{-1}$,
for an expansion age $R/(dR/dt)$ of $400 \pm 40$ years and an
expansion velocity of about 1000 km s$^{-1}$.  We suggest that the PWN
is expanding into an asymmetric nickel bubble in a conventional Type
IIP supernova.  Some acceleration of the PWN expansion is likely,
giving a true age of $480 \pm 50$ years. The pulsar's birth luminosity
was larger than the current value by a factor of 3 -- 8, while the
initial period was within a factor of 2 of its current value. We
confirm directly that Kes 75 contains the youngest known PWN, and
hence youngest known pulsar.  The pulsar PSR J1846$-$0258 has a
spindown-inferred magnetic field of $5 \times 10^{13}$ G; in 2006 it
emitted five magnetar-like short X-ray bursts, but its spindown
luminosity has not changed significantly.  However, the flux of the
PWN has decreased by about 10\% between 2009 and 2016, almost entirely
in the northern half.  A bright knot has declined by 30\% since 2006.
During this time, the photon indices of the power-law models did not
change. This flux change is too rapid to be due to normal PWN
evolution in one-zone models.

\end{abstract}


\section{Introduction}
\label{intro}

Pulsar-wind nebulae (PWNe) provide essential information on various
astrophysical phenomena.  As pulsar calorimeters, they document the
total energy injected by the pulsars, independent of beaming.  The
relativistic-wind termination shocks through which energetic particles
enter the PWNe can serve as nearby laboratories in which to study
particle acceleration in relativistic winds and jets such as those
seen in active galactic nuclei and gamma-ray burst sources.  Young
PWNe, still inside their natal shell supernova remnant (SNR), interact
with the innermost ejecta and can provide information about that
material, otherwise inaccessible.  The youngest PWNe also give
information on the youngest pulsars, whose behavior may differ from
that of the more typical pulsars that have long outlasted or escaped
from their SNRs.

While most supernovae (SNe) should result from core-collapse (CC) events, and
almost all CCSNe need to produce pulsars to produce the present-day
Galactic pulsar population \citep{lorimer06}, it is a bit surprising
that of four confirmed historical SNRs (Kepler 1604 CE, Tycho 1572,
Crab 1054, and SN 1006) and two more expansion-confirmed SNRs
(G1.9+0.3, ca.~1900, and Cas A, ca.~1680), none is a composite remnant
(shell + PWN). (3C 58 is a Crab-like PWN with no obvious shell, and in
any case is unlikely to be the remnant of an event in 1181 CE; e.g.,
\citealt{chevalier05}).  Now young pulsars can manifest themselves
through short spindown ages $P/(n-1){\dot P}$, which are upper limits
to the true ages under the assumption of normal dipole spindown with
constant braking index $n$.  One such object, G11.2$-$0.3, a shell
remnant containing an observed pulsar and PWN, was recently found to
have an age between 1400 and 2400 years, based on observations with
{\sl Chandra} between 2000 and 2013, though it cannot have resulted
from a possible supernova in 386 CE \citep{borkowski16}.  This age is
much less than its spindown age of about 23,000 years, which must be
far larger than the true age based on several arguments including the
pulsar position at the very center \citep{kaspi01}, and requiring that
the pulsar period be essentially unchanged from birth.

One other composite remnant might conceivably have resulted from a
supernova in the last two millenia: Kes 75 (G29.7$-$0.3) \citep[][see
  Figure~1]{becker76,becker83}.  An earlier distance estimate of 19 kpc
made Kes 75 a very large, luminous object, but subsequent H I
observations \citep{leahy08} gave a distance of 5.5 -- 5.9 kpc.  We
shall adopt a value of $5.8 \pm 0.5$ kpc based on the reanalysis of
\cite{verbiest12}.  At that distance, $1'' = 8.7 \times 10^{16}$ cm.
The remnant shows a partial shell of radius about $90''$, or about 2.5
pc, with a central nebula of distinct properties, about $25'' \times
35''$ in extent ($0.70 \times 0.99$ pc).  The complete absence of detectable
shell emission to the east indicates a very strong density gradient in
the medium into which Kes 75 is expanding.



The central component was shown to have a flat radio spectrum with
substantial polarization: a typical radio PWN \citep{becker76}. It was
presumed to be powered by a pulsar, but the pulsar was not discovered
until 2000, in X-rays with {\sl RXTE} \citep{gotthelf00}.  \citep[It
  still has not been detected in radio;][]{archibald08}.  At that
time, the pulsar, PSR J1846$-$0258, was found to have a period $P$ of 326
ms, a remarkably high spindown luminosity of $8 \times 10^{36}$ erg
s$^{-1}$, and a high magnetic field (for a rotation-powered pulsar) of
about $5 \times 10^{13}$ G.  This magnetic-field strength is within
the range of the so-called magnetars, neutron stars with $B \gapprox
10^{13.5}$ G powered by magnetic-field decay.  The pulsar braking
index was found by \cite{livingstone06} to be $n = 2.65 \pm 0.01$,
which gave a spindown age $P/(n-1){\dot P}$ of 884 years, the smallest
known, and implying that Kes 75 is one of the youngest supernova
remnants in the Galaxy.  Kes 75 has also been detected between 20 and
200 keV with INTEGRAL \citep{mcbride08} and between 0.3 and 5 TeV with
HESS \citep{terrier08}, though neither instrument can distinguish
between emission from the shell and from the PWN.

The pulsar in Kes 75 has proved to be highly unusual in several
respects beyond its high magnetic field.  In 2006 (just seven days
before a long {\sl Chandra} observation), the pulsar emitted a series
of five magnetar-like short X-ray bursts \citep{gavriil08}, with a
concomitant increase in the pulsar luminosity by about a factor of 6,
along with spectral softening \citep{ng08}.  It was later shown that
the spindown properties of the pulsar had changed: evidently a glitch
occurred sometime between 2005 (when the dataset fixing the earlier
braking index ended) and 2008, when a new set of phase-coherent
observations began \citep{archibald15}.  Presumably the glitch was
coincident with the X-ray bursts, though the observations do not
demand this.  The braking index was found to have decreased to $2.19
\pm 0.03$, determined over a 7-year period.  The new spindown age is
now 1230 years, still among the shortest known.  However, it is
important to note that the pulsar's spindown luminosity has not
changed significantly from its pre-flare value.  A change in $n$ of
this size, for one of the few pulsars for which timing data allow a
determination of the second period derivative, is unprecedented.

Comparison of observations of the PWN with {\sl Chandra} in 2000 and
2006 showed significant changes in the small-scale structure of the
PWN, with apparent motion of one feature giving a speed of $0.03c$
\citep{ng08}.  A jet-torus structure was identified, as often seen in
young PWNe.  \cite{ng08} performed detailed spatial analysis of the
Kes 75 PWN, fitting individual spectra to 14 separate regions, and
carefully comparing the 2006 observation to that from 2000.  They
found typical spectral behavior: a fairly hard spectrum (photon index
$\Gamma = 1.9$ where $F_\nu \propto E^{-\Gamma}$) near the pulsar,
with the spectrum softening with distance from the pulsar, although
the very hardest spectrum ($\Gamma = 1.5$) was found not in the
immediate neighborhood of the pulsar but a few arcsec away. They found
no significant changes in $\Gamma$ for the subregions between 2000 and
2006.  Another observation \citep{livingstone11} was performed in 2009
using 1/8 subarray mode, allowing a readout time of only 0.4 s to
ameliorate pileup (but as a result, mostly restricted to the pulsar/PWN
region).  These authors reported an integrated flux of the PWN
between 0.5 and 10 keV consistent with that from 2000, in spite of the
remarkable pulsar outbursts and longer-term change in properties in
2006.

Several theoretical models for the evolution of Kes 75 have appeared
recently, aiming to explain the full radio-to-TeV spectral energy
distribution.  \cite{bucciantini11} describe an evolutionary one-zone
model, for which the data are quite constraining.  They require an
additional seed photon field to account for the TeV emission, and a
very high efficiency of injection of particles.  They infer a current
mean magnetic-field strength of 20 -- 30 $\mu$G.  \cite{gelfand14}
describe a similar model, based on the pre-flare pulsar properties,
and infer an age of about 420 years.  One justification for such simple
models in the face of obvious highly complex and inhomogeneous
observed PWNe has been that the relativistic sound speed in a PWN,
$c/\sqrt{3},$ is so high that the pulsar bubbles should be essentially
isobaric.  Overall, one-zone
(``zero-dimensional'') models do a surprisingly effective job of
describing the broadband spectral-energy distribution of PWNe.

One-zone models also predict the gradual evolution of PWN properties.
PWNe are expected to evolve in luminosity as they expand.  The
timescale for such evolution, due to adiabatic expansion losses and
the slow decline of pulsar input power, is normally the dynamical
timescale, comparable to the PWN age.  Radiative properties can change
on the timescale of energy losses on particles, much longer than the
dynamical time for radio-emitting particles but only a few years for
X-ray-synchrotron-emitting electrons and positrons in typical magnetic
fields of tens to hundreds of $\mu$G.  Simple evolutionary models
\citep[e.g.,][]{rc84,gelfand09,bucciantini11} make quantitative
predictions (explicitly or implicitly) for the time-dependence of the
luminosity in different wavelength regimes, but these predictions have
rarely been tested.  One exception is the radio flux of the Crab
Nebula, predicted by \citet[][hereafter RC84]{rc84} to be declining at
radio wavelengths at $(0.26 \pm 0.1)$\% yr$^{-1}$, a prediction
verified by \cite{aller85}.  These are global predictions for
integrated fluxes, and expected rates of change are slow.  Small-scale
brightness changes, on the other hand, have been followed in a few
well-known cases, including the Crab wisps and X-ray knot
\citep{hester08} and the Vela jet \citep{durant13}, and several other
PWNe including Kes 75 show motions of small features \citep{ng08}.

One-zone models also assume the applicability of the simplest
dipole-spindown models for pulsars. For normal magnetic-dipole
spindown with constant $n$, the spindown age $t_{\rm sd} \equiv P/(n -
1)\dot P$ increases with time: $t_{\rm sd} = \tau + t$ where $\tau
\equiv t_{\rm sd}(0)$ is the spindown time at birth.  Then the pulsar
luminosity $L(t)$ decays as
\begin{equation} 
L(t) = \frac{L_0}{(1 + t/\tau)^p} \ \ {\rm where} \ \ p \equiv \frac{n+1}{n-1}.
\label{lum}
\end{equation}
The pulsar period $P$ then obeys $P = P_0
\left(L_0/L\right)^{1/{n+1}}$. However, for pure magnetic-dipole
field configurations, the braking index $n = 3$, which is never
observed among the dozen or so pulsars with measured braking indices
\citep{espinoza16}.  More complex field geometries, and other
modifications of the simple picture, have been proposed to account for
observed values of $n$ \citep[e.g.,][]{gao17, akgun17, antonopoulou18}; 
these models produce different spindown histories than simple dipole
spindown.

For Kes 75, the significant changes in pulsar spindown properties
observed since 2006 cast an additional shadow of uncertainty over the
age estimates based on simple dipole spindown.  But if Kes 75 is really less
than 1000 years old, and at a distance of only about 6 kpc, its
expansion should be measurable.
On this basis, we obtained a 150 ks {\sl Chandra} exposure in 2016.

\section{Observations}
\label{obssec}

\begin{deluxetable}{lccc}
\tablecolumns{4}
\tablecaption{{\sl Chandra} Observations of Kes 75 \label{observationlog}}
\tablehead{
& & & Effective Expo- \\
\colhead{Date} & Observation ID & Roll Angle & sure Time \\
& & (deg) & (ks) }

\startdata
2000 Oct 15--16   &   748 & 279 & 31.68 \\
2006 Jun 05       &  7337 & 133 & 17.36 \\
2006 Jun 07--08   &  6686 & 133 & 49.02 \\
2006 Jun 09       &  7338 & 133 & 39.25 \\
2006 Jun 12--13   &  7339 & 133 & 44.05 \\
2009 Aug 10--11   & 10938 & 249 & 44.25 \\
2016 Jun 08--09   & 18030 & 133 & 84.76 \\
2016 Jun 11--12   & 18866 & 133 & 60.99 \\
\enddata

\end{deluxetable}

{\sl Chandra} observed Kes 75 at three epochs during its first decade
of operations: in 2000 (Epoch I), 2006 (Epoch II), and 2009 (Epoch
III). The most recent Epoch IV observation with {\sl Chandra} took
place in 2016 June in two separate pointings
(Table~\ref{observationlog}), with the remnant again placed on the
Advanced CCD Imaging Spectrometer (ACIS) S3 chip.  Very Faint mode was
used in order to reduce the particle background for this low surface
brightness target. We used CIAO version 4.9 and CALDB version 4.7.4 to
reprocess these Epoch IV observations. The bright central pulsar
\psr\ was used to align the 2016 June 08--09 and 11--12
observations. After screening for particle flares, the total effective
exposure time is 146 ks.

Epoch I--III observations (Table~\ref{observationlog}) were
reprocessed as for Epoch IV, but the particle background rate is
higher at Epochs I and III because Faint mode was used instead of Very
Faint mode for these relatively shallow (32 and 44 ks)
observations. For Epoch II, the total effective exposure is 150 ks,
comparable in length to Epoch IV. Three shorter pointings from 2006
(observation IDs 7337 -- 7339) were aligned to the longest 2006
pointing (observation ID 6686).

The pulsar \psr\ was used for the inter-epoch alignment. It is by far
the brightest point source in the {\sl Chandra} field of view, located
close ($<1 \farcm 1$) to the optical axis, so its position can be
determined with high precision for each observation listed in
Table~\ref{observationlog}. In order to measure its position, we
applied the CIAO task {\tt srcextent} to data processed with Faint
(instead of Very Faint) mode as appropriate for bright point
sources. The pulsar's point spread function (PSF) is approximated by a
2D Gaussian in {\tt srcextent}, allowing us to find its centroid and
width after fitting this Gaussian to the data. The
estimated\footnote{See equation~(11) in Houck, J.~C. 2007,
  http://cxc.harvard.edu/csc/\\ memos/files/Houck\textunderscore
  source\textunderscore extent.pdf.} positional uncertainties do not
exceed 22 mas (at $90\%$ confidence level). This corresponds to
$1\sigma$ relative errors of $<0.1\%$ at a radial distance of
$15\arcsec$ away from the pulsar. As this is much smaller than the
statistical errors of our expansion measurements, alignment errors
relative to the pulsar's reference frame can be safely ignored.

Although the pulsar's reference frame is most appropriate for
measuring the expansion of the PWN, there is a possibility of a
substantial (several hundred km s$^{-1}$) pulsar kick. In the
framework of freely expanding uniform ejecta discussed in
\S~\ref{discussion}, this would have resulted in a nonnegligible net
motion of the entire PWN/PSR system relative to the SN and its local
frame of reference, but otherwise without any influence on the PWN
dynamics.  If large enough, the tangential component of this motion
would manifest itself as a measurable proper motion of the pulsar
relative to background and foreground point sources. Then, these
sources would appear misaligned in the pulsar's reference frame,
particularly between Epochs I and IV.

We examined the relative positions of sufficiently bright point
sources between Epochs I and IV (after alignment of observations to
the pulsar reference frame). The relatively short (32 ks) duration of
the Epoch I observation limits the number of matching sources suitable
for reasonably accurate measurements to 10. They are rather faint on
average, with a median number of counts of only 19 at Epoch I. Their
off-axis angles range from $1\farcm 2$ to $3\farcm 8$.  Their
positions were found using the CIAO task {\tt wavdetect}, while
$1\sigma$ positional errors were estimated using equation~(14) of
\citet{kim07}.  These positions differ by an average of $ \Delta
\alpha \cos \delta = -120$ mas and $ \Delta \delta = -30$ mas between
Epochs I and IV, but there is a large scatter ($210$ mas and $170$ mas,
respectively) around these values. The measured point source
displacements range from 95 to 680 mas, with the median of 160
mas. After their normalization by $1\sigma$ errors, they range from
0.27 to 2.2.  Their distribution is well described by the Rayleigh
distribution with a scale of $0.83 \pm 0.13$ (estimated using the
method of maximum likelihood) that is statistically consistent with
unity. Therefore, we find no evidence for misalignment of point
sources or a discernible pulsar motion. This conclusion must be
considered as tentative because our {\tt wavdetect}-derived positions
do not rely on realistic models of the {\sl Chandra} PSF.


\begin{figure}
\hspace*{-5mm}
\epsscale{1.28}
\plotone{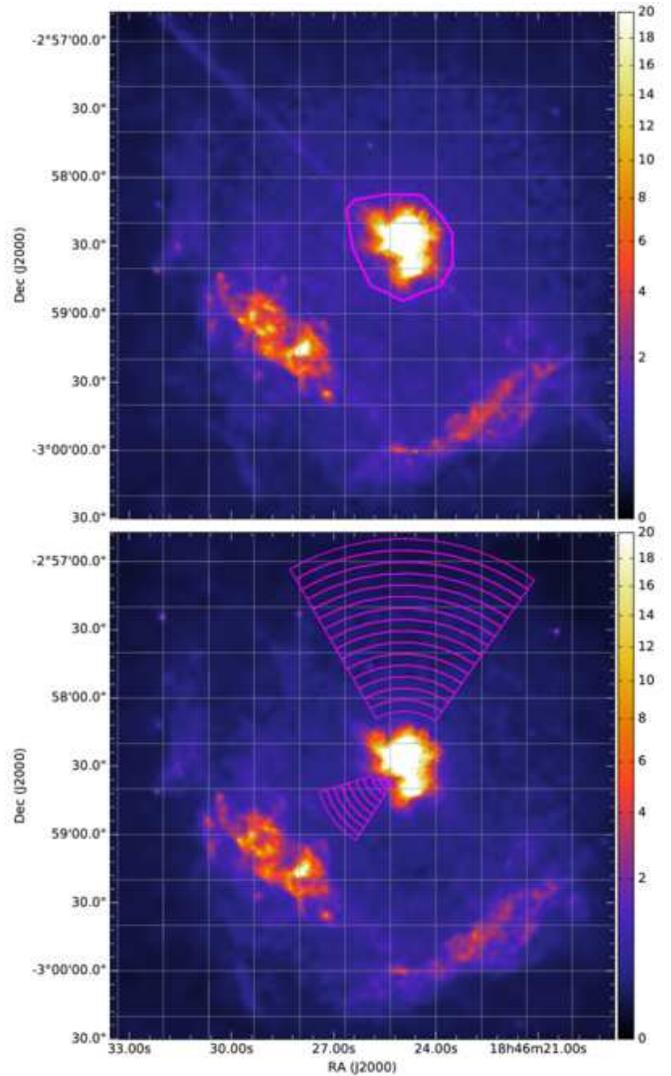}
\caption{ X-ray images of Kes 75 in the $0.7$--$8$ keV energy range
  from 2006 (top) and 2016 (bottom), smoothed with the multiscale
  partitioning method of \citet{krishnamurthy10}. The bright
  pulsar wind nebula at its center is saturated in order to show the
much fainter shell structure. The 2006 image also
  shows the region used for flux extraction of the entire PWN.  A
  dust-scattered halo is apparent in 2006. Its radial profile (see
  Figure \ref{fig-halo}) was measured within overlaid regions on the
  2016 image. The scales are in counts per $0 \farcs 432 \times 0
  \farcs 432$ image pixel.}
\label{xrays06and16}
\end{figure}

Epoch II and IV images of Kes 75, extracted from merged and smoothed
{\sl Chandra} data cubes, are shown in
Figure~\ref{xrays06and16}. Expansion of the shell is subtle but
discernible by eye with the help of the coordinate grid, and its
complex motion will be described in a separate
investigation. Out-of-time events from the pulsar can be seen as
diagonal ``streaks'' in these images, being much more prominent in
2006 because of the much brighter pulsar at this epoch. Excess
emission in the interior of the remnant is also apparent at this
epoch, presumably a halo produced by scattering of the pulsar's X-rays
by interstellar dust present along the line of sight to Kes 75. This
excess emission had been noticed previously \citep{ng08}, but now we
can examine it in more detail by comparing the images shown in
Figure~\ref{xrays06and16}. This contaminating halo emission must be
taken into account when measuring expansion of the PWN.

The XSPEC spectral analysis package \citep{arnaud96} was used to
examine X-ray spectra, which were extracted from individual
observations and added together to obtain merged spectra. (The
response files for each epoch's obsID's were averaged). Spectra of Kes
75 PWN were modeled with an absorbed power law, using the solar
abundances of \citet{grevesse98} in the {\tt phabs} absorption model.
In order to preserve the Poisson nature of the statistics, we modeled
rather than subtracted background for spectral fitting.

\section{Expansion of the PWN}

The time baselines between Epochs I $-$ III and Epoch IV range from
6.83 to 15.65 years, long enough to reliably measure expansion of the
PWN. We use a variation of the method described by us previously in
our studies of the youngest Galactic SNR G1.9+0.3
\citep{carlton11,borkowski14} and young CC SNRs G11.2$-$0.3 and Kes 73
\citep{borkowski16,borkowski17}. First, we extracted two data cubes
from the merged Epoch II and the merged Epoch IV observations, with
$300^2$ image pixels and 16 spectral channels, in the energy range
from 0.7 to 8 keV, encompassing the entire PWN. The spatial pixel size
is $0 \farcs 216 \times 0 \farcs 216$.  We then removed the bright
pulsar from these data cubes by masking it with a circle $2 \farcs 4$
in diameter. In each spectral channel, we replaced pixel values within
this circle by simulated values assuming that the mean surface
brightness there is constant and equal to the mean surface brightness
within an ellipse centered on the pulsar, $5 \farcs 4 \times 2 \farcs
7$ in size and with its long axis perpendicular to the PWN
jet. Poisson statistics were assumed in these simulations. We rebinned
these filtered data cubes by a factor of 2 in the spatial dimension,
giving us final data cubes, $150^2 \times 16$ in size. The final
spatial pixel size is $0 \farcs 432 \times 0 \farcs 432$ (slightly
less than an ACIS $0 \farcs 492$ pixel). We smoothed these data cubes
with the non-local PCA method of \citet{salmon14}. This method
combines elements of dictionary learning and sparse patch-based
representation of images (or spectral data cubes) for photon-limited
data. Because this Poisson-PCA method is computationally intensive,
relatively small ($150^2 \times 16$) data cubes, heavily binned along
the spectral dimension as described above, were smoothed using patches
$5^2 \times 6$ in size. The moderate spatial patch size of $2 \farcs
16 \times 2 \farcs 16$ preserves sharp spatial structures seen in the
bright jets and in much fainter filamentary features found along the
periphery of the PWN, while a large patch size in the spectral
dimension is suitable for the synchrotron-dominated spectra of the PWN
that vary smoothly across the entire spectral range of {\sl
  Chandra}. With the patch size chosen, the most important parameters
that control the smoothing of data cubes are the order $l$
of the Poisson-PCA method, and the
number of clusters $K$ into which patches are grouped prior to
estimation of intensities. We used $l=6$ and $K=30$ for the 2006 and
2016 data cubes of the Kes 75 PWN.

\begin{figure}
\hspace*{-5mm}
\epsscale{1.3}
\plotone{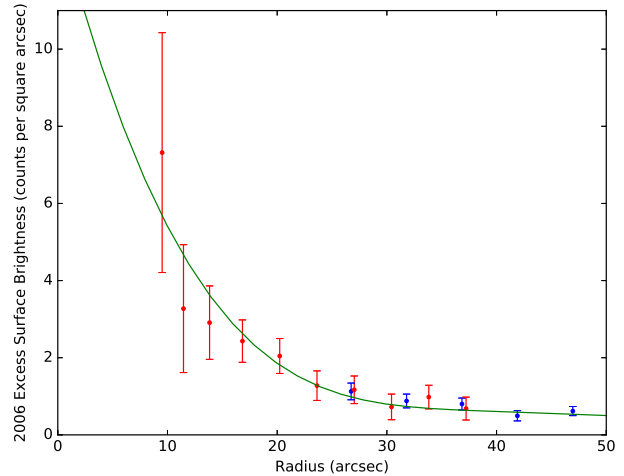}
\caption{Radial profile of the 2006 X-ray halo at distances $<50''$ away
  from the pulsar, for northern (in blue) and
  southeastern (in red) regions shown in Figure \ref{xrays06and16}.
  Smoothed profile is in green. }
\label{fig-halo}
\end{figure}

Images extracted from the smoothed data cubes must be corrected for effects of
the time-varying background. We account for the time-varying particle
background by determining the combined X-ray and particle background in
a source-free region on the ACIS S3 chip, and then subtracting it from the
smoothed images.
However, the spatially-varying X-ray halo seen in the 2006 image (top panel in
Figure~\ref{xrays06and16}) contributes most to the temporal background
variations in the vicinity of the PWN. Within each of the concentric regions
shown in the lower panel of Figure~\ref{xrays06and16}, we
determined this halo contribution by subtracting the (exposure-weighted) 2016
image from the 2006 image. All these regions are centered on
the pulsar. They are located outside of the PWN, including the innermost region
only $10''$ southeast of the pulsar. If possible, they have been chosen not to
overlap with the SNR shell emission seen in projection toward the center of the
remnant.
The measured radial surface brightness profile of the halo at distances 
$< 50''$ away from the pulsar is shown in 
Figure~\ref{fig-halo}. A smoothed profile is also shown.
The halo 
surface brightness north and southeast of the pulsar is well matched where they
overlap in radius, fully consistent with a spherically-symmetric halo centered
on the pulsar. Its brightness steeply increases toward the pulsar, suggesting
that scattering of X-rays by the interstellar dust located between us and the
pulsar is responsible for this transient halo. We use the 
smoothed profile shown in Figure~\ref{fig-halo} to model the halo contribution
to the background at Epoch II. Since it is difficult to measure
the halo surface brightness profile close to the pulsar, this contribution
must be considered as very
uncertain at small ($<10''$) distances away from the pulsar. 

\begin{figure}
\epsscale{1.08}
\plotone{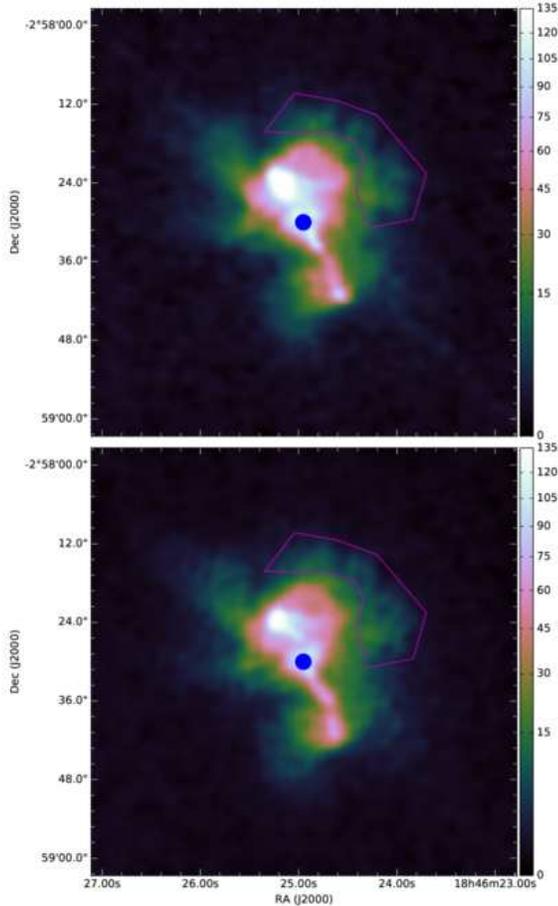}
\caption{X-ray images of Kes 75 PWN in the $0.7$--$8$ keV energy range
  from 2006 (top) and 2016 (bottom), smoothed with the non-local PCA
  method of \citet{salmon14}. Background has been subtracted as
  described in the text. The pulsar has been masked out. Expansion of
  the PWN was measured along its northwest edge (within the region
  shown in magenta). Intensities are shown with the cubehelix color scheme
  of \citet{green11}.  The scale is in counts per $0 \farcs 432 \times 0
  \farcs 432$ image pixel.}
\label{fig-pwn06and16}
\end{figure}


\begin{deluxetable*}{lcccc}
\tablecolumns{5}
\tablewidth{0pc}
\tablecaption{Expansion of Kes 75 Pulsar Wind Nebula}

\tablehead{
\colhead{Baseline}  & $\Delta t$\tablenotemark{a} & $S$\tablenotemark{b} & Expansion\tablenotemark{c}  & Expansion Rate\tablenotemark{c} \\
\colhead{} & (year) & & (\%) & (\%\ yr$^{-1}$)}

\startdata
2000 -- 2016 & 15.65 & $1.222 \pm 0.047$ & ($4.27 \pm 0.61$)& ($0.273 \pm 0.039$) \\
& 15.65 & \nodata & $4.16 \pm 0.72$ & $0.266 \pm 0.046$ \\
2006 -- 2016 & 10.00 & $1.148 \pm 0.022$ & ($2.38 \pm 0.33$)& ($0.238 \pm 0.033$) \\
& 10.00 & ($0.857 \pm 0.017$)\tablenotemark{d} &($2.17 \pm 0.38$)\tablenotemark{d} & ($0.216 \pm 0.038$)\tablenotemark{d} \\
& 10.00 & \nodata &$2.27 \pm 0.51$ & $0.227 \pm 0.051$ \\
2009 -- 2016 & 6.83 & $1.134 \pm 0.037$ & ($1.81 \pm 0.54$)& ($0.265 \pm 0.079$) \\
& 6.83 & \nodata &$1.70 \pm 0.66$ & $0.249 \pm 0.097$ \\
(2000+2006+2009) -- 2016 & \nodata & \nodata & \nodata & $0.249 \pm 0.023$\\
\enddata

\tablecomments{ Measurements were obtained in the region shown in Figure 3.  All errors are $1\sigma$.
For each time baseline, the different
lines differ in whether systematic effects were taken into account; values in
parentheses are before correction for systematic effects.
}
\tablenotetext{a}{Baseline length.}
\tablenotetext{b}{Model surface brightness scaling.}
\tablenotetext{c}{Values in brackets are before correction for systematic effects.}
\tablenotetext{d}{Model derived from the 2006 (instead of 2016) data.}
\label{expansiontable}
\end{deluxetable*}

Images of the PWN extracted from the smoothed 2006 and 2016 data cubes, after
background subtraction (including the halo contribution discussed above), are
shown in Figure~\ref{fig-pwn06and16}. These images, after normalization by
monochromatic ($E=3$ keV) exposure maps, are used as models for the
complex spatial brightness distribution of the PWN. These models are fit to
data consisting of unsmoothed images using the maximum likelihood method of
\citet{cash79} as appropriate for data dominated by Poisson statistics. In
these fits, we allow for change in the physical image scale
and in the surface brightness scale factor $S$. Expansion is centered on the
pulsar.
Except for Epoch II, a uniform background is assumed, with its value
determined in a source-free region on the ACIS S3 chip. For Epoch II, we add
the X-ray halo to the background model, using the smoothed halo profile shown
in Figure~\ref{fig-halo}. Spatial variations in the effective exposure time are
accounted for with help of the monochromatic ($E=3$ keV) exposure maps.

The PWN is dominated by the prominent northern and southern jets whose
morphologies have changed greatly between 2006 and 2016
(Figure~\ref{fig-pwn06and16}). Near these jets, it is difficult to separate
relatively slow motions expected from the PWN expansion from
much more pronounced rapid changes caused by the
short-term pulsar activity. Expansion of the PWN should be most easily 
detected far away from the jets where fast morphological and brightness
variations are not expected. The relatively faint northwest rim of the PWN
(see Figure~\ref{fig-pwn06and16}) is the most suitable region for measuring
expansion of the PWN as it occupies a substantial (over $90{\arcdeg}$)
range in azimuth, and its irregular edge is relatively sharp.

We first measured expansion of the northwest rim of the PWN between
Epochs II and IV using two image pairs: (1) smoothed 2016 and
unsmoothed 2006 images, and (2) smoothed 2006 and unsmoothed 2016
images. Results are listed in the third and fourth rows of
Table~\ref{expansiontable}. The measured expansion is $2.38\% \pm
0.33\%$ and $2.17\% \pm 0.38\%$, respectively. Since these
measurements are not independent, the small ($0.2\%$) difference
between them is caused by a bias inherent in our expansion measurement
method. We attribute this systematic effect to smoothing that
artificially makes the PWN slightly larger, leading to an
overestimation of expansion for the image pair (1) and its
underestimation for the image pair (2). Since exposure times are
comparable for Epochs II and IV, this bias can be removed by averaging
the measured expansions. The bias magnitude is $0.11\%$ for each image
pair, with the bias positive for the image pair (1) and negative for
the image pair (2). The averaged expansion is $2.27\% \pm 0.51\%$
(errors have been added in quadrature to account for uncertainties in
the smoothed images arising from photon noise). With the time baseline
of 10 years, this corresponds to an expansion rate of $0.227\% \pm
0.051\%$ yr$^{-1}$.

The long (15.65 years) time baseline between Epochs I and IV allows
for an independent and reliable measurement of expansion of the PWN
northwest rim, although the short (32 ks) exposure time at Epoch I
limits its accuracy. We used the smoothed 2016 image in combination
with an unsmoothed 2000 image to arrive at an expansion of $4.27\% \pm
0.61\%$ (see the first row of Table~\ref{expansiontable}). After
reduction by $0.11\%$ due to the bias caused by smoothing, the expansion
becomes $4.16\% \pm 0.72\%$ (the error increased modestly as we
combined in quadrature the statistical errors arising from photon
noise for both epochs).
This unbiased expansion
measurement is listed in the second row of Table~\ref{expansiontable}. The
corresponding expansion rate, also listed in this Table, is
$0.266\% \pm 0.046\%$ yr$^{-1}$, in good agreement with the expansion rate of
$0.227\% \pm 0.051\%$ yr$^{-1}$ measured between Epochs II and IV. 

We also measured expansion between Epochs III and IV in the same way as for
Epochs I and IV (results are also listed in Table~\ref{expansiontable}).
The unbiased expansion rate is $0.249\% \pm 0.097\%$ yr$^{-1}$. The best
estimate of the expansion rate, $0.249\% \pm 0.023\%$ yr$^{-1}$, is obtained by
combining all three independent expansion rate measurements. The variance of 
these measurements is small, as reflected by the small error of
$0.023\%$ yr$^{-1}$ for this averaged rate.

If the expansion is spatially uniform within the region shown in
  Figure 3 within which expansion is measured, the corresponding
  spatial velocity is proportional to distance from the pulsar.  The
  irregular outer edge of the PWN in that region is at a radius of
  about $18''$, giving, for our adopted distance of 5.8 kpc, a
  velocity of 1200 km s$^{-1}$.  Smaller radii then have
  proportionally smaller velocities.  In Section 5 we adopt 1000 km
  s$^{-1}$ as an estimate of the PWN expansion velocity.

There is a significant decrease in the surface brightness in 2016 for
the PWN northwest rim (Table~\ref{expansiontable}). When using the
smoothed 2016 image as a model, the surface brightness scaling factor
$S$ varies between $1.13$ and $1.22$, systematically increasing with
the measured expansion. In the absence of intrinsic flux variations,
we expect $S$ to be larger than unity, with $S-1$ being twice as large
as the measured expansion up to moderate (several percent) expansion
values \citep{carlton11}. So we expect $S$ to range from $1.04$ for
the Epoch III and IV image pair to $1.09$ for the Epoch I and IV image
pair.  A decrease in flux of about $10\%$ in 2016 is required to
explain the larger than expected $S$ obtained while fitting for
expansion. For comparison, we estimate that the 2006 X-ray halo
contribution to the measured flux of the PWN northwest rim is only a
few percent. So this flux decrease is highly significant and quite
surprising.

\section{Flux and morphology variations}

\begin{figure}
\centerline{\includegraphics[width=7.5cm]{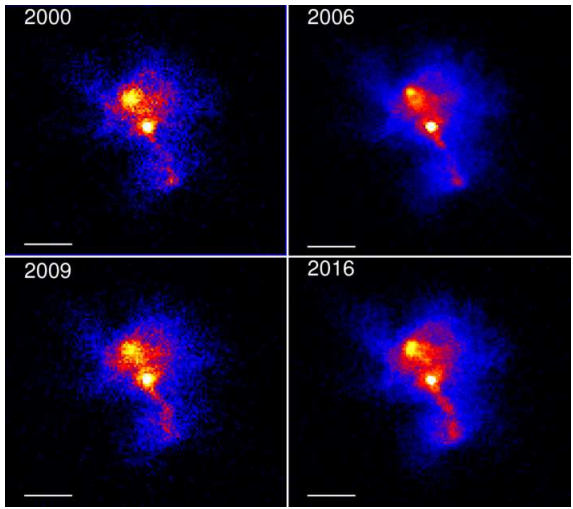}}
\caption{Images of the Kes 75 PWN at four epochs.  No background has
been subtracted.  The white bar indicates
$10''$.  
}
\label{fig-4ims}
\end{figure}

As was noted by \cite{ng08} and \cite{livingstone11}, small-scale
morphological changes occurred between 2000, 2006, and 2009.  This has
continued with the 2016 observations (Figure~\ref{fig-4ims}).  The
relative brightness of the northern knot (Figure~\ref{fig-regions}) has
varied significantly (see below), but there are no major changes.  The
ends of the jets appear not to move out significantly, though the
southern jet end may move transversely slightly, and subtle outward
motions of a feature just south of the pulsar were reported by
\cite{ng08}.  \cite{livingstone11} show profiles along the jet,
illustrating the relatively small changes from 2000 to 2009.

The X-ray bursts and glitch may have caused changes in the PWN flux,
though neither \cite{ng08} nor \cite{kumar08} found a significant
change between 2000 and 2006.  The pulsar was much brighter in 2006
than previously, by about a factor of 6 \citep{ng08, kumar08},
contaminating the flux of the PWN with the trail of out-of-time events
from the pulsar.  More importantly, the excess scattered light we detect in
the 2006 observation will contribute to the measured PWN flux.  The
model of Section 3, integrated over the region shown in
Figure~\ref{xrays06and16}, accounts for about 5\% of the total counts
in the PWN for that epoch.  



\begin{figure}
\centerline{\includegraphics[width=8cm]{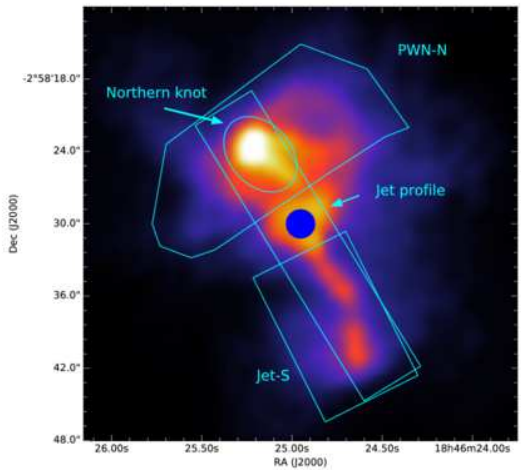}}
\caption{Regions used for subsequent analysis, superposed on smoothed
  2016 image with pulsar removed.  Long rectangle: region for jet
  profiles.  Small ellipse: Northern knot.  Northern cyan polygon: region
  PWN-N.  Southern rectangle: region Jet-S.  }
\label{fig-regions}
\end{figure}

\begin{figure}
\centerline{\includegraphics[width=8.5cm]{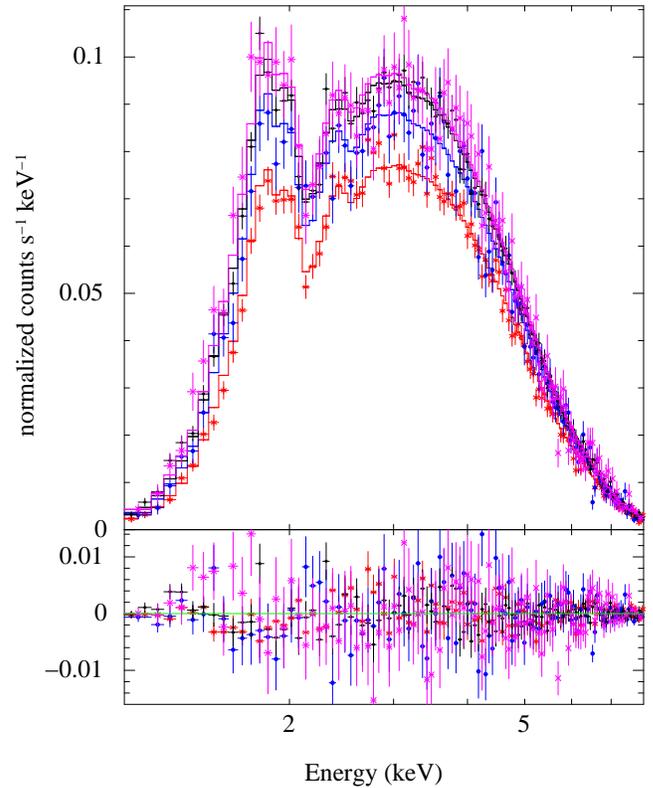}}
\caption{Spectra from four epochs for region PWN-N (the bulk of the
  northern nebula).  From top down, data are from 2000, 2006, 2009,
  and 2016. Data have been binned by a factor of four, then adaptively
  binned, for display only.  The models shown were fit as described in
  the text.  The decrease with time is obvious.}
\label{fig-spectra}
\end{figure}

\begin{figure}
\centerline{\includegraphics[width=8cm]{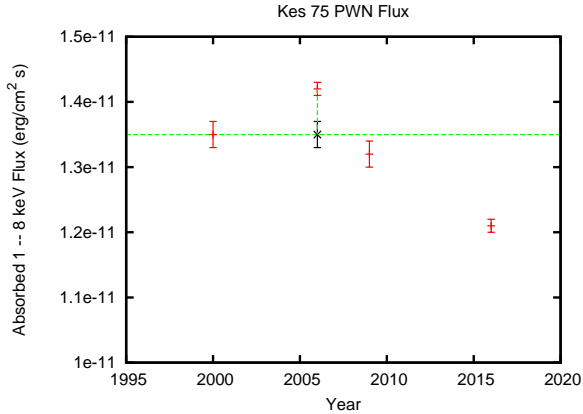}}
\caption{Total PWN flux, not corrected for absorption, excluding the
  pulsar (1 -- 8 keV).  The flux in 2006 is shown both before (red)
  and after (black) subtraction of the estimated scattering
  contribution.  The horizontal line is the 2000 value.}
\label{fig-PWNtot}
\end{figure}

The results of Section 3 indicate that the northwest rim of the PWN
appears to have decreased in flux by about 10\% between 2006 and 2016.
Even taking into account a few percent excess flux in 2006 from the
scattering halo, the decrease is significant and motivated a deeper
investigation.  To search for more extensive changes, we excluded a
circular region $3''$ in diameter around the pulsar position from the
region shown in the upper panel of Figure~\ref{xrays06and16}, and fit
the PWN with a power-law with absorption, constraining the total
(absorbed) flux rather than the normalization at some energy (XSPEC
task {\tt cflux}).  This method gives absorbed fluxes that are
relatively insensitive to the amount of absorption.  To minimize
  effects of uncertainties in absorption, however, we fit spectra from
  all four epochs jointly, with absorbing column densities tied
  together. A similar method was used to measure fluxes
  in the three smaller regions shown in Figure~\ref{fig-regions}.
Figure~\ref{fig-spectra} shows the spectra from all four
epochs for region PWN-N, along with the model fits.  The flux decrease
is apparent.  The results for integrated fluxes between 1 and 8 keV
are shown in Table~\ref{fluxtable} and plotted in
Figure~\ref{fig-PWNtot}.  Errors given there are statistical
  only; systematic calibration errors can reach 3\%
\footnote{http://cxc.harvard.edu/cal/summary/Calibration\_Status\_Report.html\#ACIS\_EA}.  Thus changes above about 5\% are significant.


\begin{figure}
\centerline{\includegraphics[width=8cm]{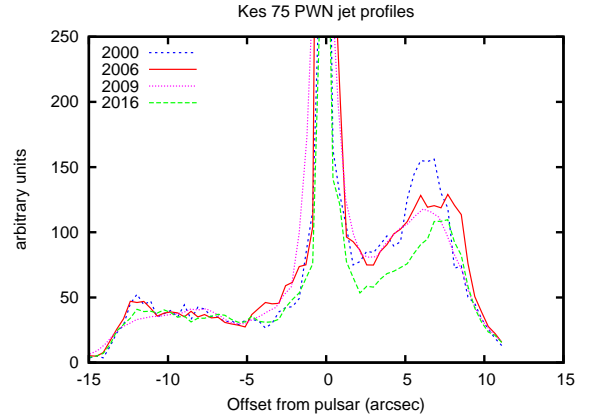}}
\caption{Profiles of the jet at four epochs, using region shown in
  Figure~\ref{fig-regions}.  Positive offsets are in the northern
  direction.  }
\label{fig-profiles}
\end{figure}

\begin{figure}
\centerline{\includegraphics[width=8cm]{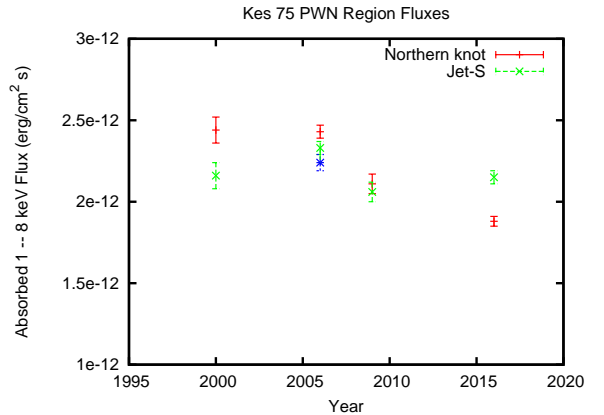}}
\caption{Fluxes of regions Jet-S (in green) and the northern knot (in
  red) (see Figure~\ref{fig-regions}) as a function of time.  The 2006
observation for Region Jet-S shows the flux corrected for scattered
light (reduced by 4\%); the 1\% correction for the northern knot is 
within statistical measurement errors.}
\label{fig-jet2fluxes}
\end{figure}

We find that the PWN overall decreased in integrated flux between 2006
and 2016 by $(17 \pm 1)$\%, a highly significant change.  If the 5\%
contribution in total counts in 2006 due to scattering produces a
comparable contribution in total flux (i.e., neglecting differences in
the spectrum of scattered pulsar X-rays and that of the PWN), the
change is still of order 11\%.  In addition, the decrease from the
2000 value, $(12 \pm 2)$\%, is also significant, as is the decrease
from 2009, $(9 \pm 2)$\%.  However, the decrease is not spatially
uniform.  Jet profiles in Figure~\ref{fig-profiles} show that the
southern jet has remained roughly constant, while the northern one has
experienced changes in morphology and overall brightness.
Figure~\ref{fig-jet2fluxes} quantifies this; the large northern
portion of the PWN has faded by $(20 \pm 2)$\% since 2000, with much
of that decrease attributable to the bright northern knot which
declined by $(30 \pm 4)$\%.  The contribution of scattered light
  in the 2006 observation to total PWN flux is about 5\% as mentioned
  above; in addition, scattered light contributes about 4\% to region
  Jet-S, and about 1\% to the northern knot. Figures~\ref{fig-PWNtot}
and~\ref{fig-jet2fluxes}
include the corrections to the total nebula and to region Jet-S.

The evidence for a flux decrease is unambiguous.  However, we made an
attempt to quantify its significance.  The two short observations, in
2000 and 2009, provide the strictest test.  First, we confront the
well-known problem of using C statistics for estimates of goodness of
fit \citep{connors07}.  Recent work \citep{kaastra17} provides some
numerical approximations to allow the calculation of expectations
$C_e$ and variances $(\delta C)^2$ of the C statistic for particular
situations.  Our spectral fits for region PWN-N give values for $(C -
C_e)/\delta C$ of $-1.0$ and 1.3 for the 2000 and 2009 fits -- quite
acceptable, and consistent with the visual impression of
Figure~\ref{fig-spectra}.

We can provide a more rigorous test for the presence of the flux
decrease using the likelihood-ratio test \citep[LRT;][]{cash79}.  Our
fitting indicates a drop in flux in Region PWN-N between 2000 and 2009
of 6.8\%.  To test the null hypothesis of no change in flux, we re-fit
with the fluxes for those two years tied together.  This produced an
increase in the C statistic of 16.5.  Using the LRT, we find that the
likelihood of the null hypothesis is $4.7 \times 10^{-5}$.  Other
similar tests provide even more extreme rejections of the null
hypothesis.  However, since our statistical errors are no larger than
the 3\% systematic calibration errors, whose distribution is unknown,
this exercise is of little quantitative value, and we have not
performed it for other spectral fits.

\begin{deluxetable}{lcccc}
\tablecolumns{5}
\tablewidth{0pc}
\tablecaption{Kes 75 Pulsar-Wind Nebula Fluxes \label{fluxtable}}
\tablehead{
\colhead{Region} & $F(2000)$ & $F(2006)$ & $F(2009)$ & $F(2016)$
}
\startdata
PWN\tablenotemark{a}    & $13.5 \pm 0.2$  & $14.2 \pm 0.1$  & $13.2 \pm 0.2$  & $12.1 \pm 0.1$ \\
PWN-N\tablenotemark{b}  & $6.41 \pm 0.13$ & $6.42 \pm 0.06$ & $6.00 \pm 0.11$ & $5.34 \pm 0.06$\\
Northern knot\tablenotemark{c} & $2.44 \pm 0.08$ & $2.43 \pm 0.04$ & $2.11 \pm 0.06$ & $1.88 \pm 0.03$\\
Jet-S\tablenotemark{d}  & $2.16 \pm 0.08$ & $2.33 \pm 0.04$ & $2.06 \pm 0.06$ & $2.15 \pm 0.04$\\
\enddata
\tablecomments{Fluxes (1 -- 8 keV) in units of $10^{-12}$ erg cm$^{-2}$ s$^{-1}$, not corrected for
absorption. Errors are 90\% confidence.}
\tablenotetext{a} {$N_H = 4.51 \pm 0.04 \times 10^{22}$ cm$^{-2}$.}
\tablenotetext{b} {$N_H = 4.48 \pm 0.06 \times 10^{22}$ cm$^{-2}$.}
\tablenotetext{c} {$N_H = 4.46 \pm 0.10 \times 10^{22}$ cm$^{-2}$.}
\tablenotetext{d} {$N_H = 4.71 \pm 0.11 \times 10^{22}$ cm$^{-2}$.}
\end{deluxetable}

\begin{deluxetable}{lcccc}
\tablecolumns{5}
\tablewidth{0pc}
\tablecaption{Kes 75 Pulsar-Wind Nebula Photon Indices \label{spixtable}}
\tablehead{
\colhead{Region} & $\Gamma (2000)$ & $\Gamma (2006)$ & $\Gamma (2009)$ & $\Gamma (2016)$
}
\startdata
PWN\tablenotemark{a}    & $2.01 \pm 0.03$ & $2.03 \pm 0.02$ & $1.97 \pm 0.03$ & $2.00 \pm 0.02$ \\
PWN-N\tablenotemark{b}  & $1.99 \pm 0.05$ & $1.97 \pm 0.03$ & $1.94 \pm 0.04$ & $1.95 \pm 0.03$\\
Northern knot\tablenotemark{c} & $1.83 \pm 0.08$ & $1.81 \pm 0.05$ & $1.82 \pm 0.07$ & $1.83 \pm 0.05$\\
Jet-S\tablenotemark{d}  & $1.89 \pm 0.08$ & $1.93 \pm 0.05$ & $1.97 \pm 0.08$ & $1.95 \pm 0.05$\\
\enddata
\tablecomments{Fits are from 1 to 8 keV.  Errors are 90\% confidence.}
\tablenotetext{a} {$N_H = 4.51 \pm 0.04 \times 10^{22}$ cm$^{-2}$.}
\tablenotetext{b} {$N_H = 4.48 \pm 0.06 \times 10^{22}$ cm$^{-2}$.}
\tablenotetext{c} {$N_H = 4.46 \pm 0.10 \times 10^{22}$ cm$^{-2}$.}
\tablenotetext{d} {$N_H = 4.71 \pm 0.11 \times 10^{22}$ cm$^{-2}$.}
\end{deluxetable}

Fitted photon indices $\Gamma$ are shown in Table~\ref{spixtable}.
When absorption values are fixed for all epochs, there are no
significant variations in photon index, even while the flux drops
substantially by 2016 for all but the Jet-S region. The values we find
tend to be steeper by about 0.1 than those reported in \cite{ng08},
probably because our choice of \citet{grevesse98} abundances produces
considerably larger absorbing column densities than those resulting
from earlier abundance sets.  However, we concur that the northern
knot has a harder spectrum than the PWN as a whole.

\section{Discussion}
\label{discussion}

\subsection{PWN Expansion}

The results of Table~\ref{expansiontable} give an expansion age
$R/(dR/dt)$ of about $400 \pm 40$ years.  If expansion had taken place at
constant speed, this would be the true age.  Since 2008, the pulsar
has a period of 328 ms and a braking index $n$ of 2.19
\citep{archibald15}, but before then, the braking index was 2.65
\citep{livingstone06}.  The dominant evolution of Kes 75 and its
pulsar has taken place with the earlier value, which with due caution
we take to have been constant since birth.  For \psr, the earlier
value of $n$ gives a decay index $p$ (Equation~\ref{lum}) of 2.21.

For $t \lapprox \tau,$ simple models
(\citetalias{rc84}; \citealt{vds01})
predict the
PWN radius to grow as $R^{6/5}$, or an expansion index $m \equiv vt/R
= 1.2$, so that the true age is 1.2 times the expansion age, or about
430 -- 530 years.  However, this gives $\tau \sim 350 - 450$ years,
respectively, and (from Equation~\ref{lum}) $L$ between 0.13 and 0.23
times $L_0$.  In this case, we might expect the expansion's
acceleration to have decreased toward $m = 1$, i.e., the calculation
is not self-consistent.  However, taking $m = 1$ similarly gives $L =
(0.2 - 0.3)L_0$, close enough to $L_0$ that less deceleration will
have occurred.  We conclude in any case that to within 50\%, $t \sim
\tau$, so we confirm that the true age of Kes 75 and its pulsar is
between 360 and 530 years -- the youngest known PWN in the Galaxy.

We performed our expansion measurement aligning the images on the
pulsar itself.  Even if the pulsar is moving with respect to the inner
unshocked SN ejecta, those ejecta are expanding with a ``Hubble law''
velocity profile, $v \propto r$, and an observer at any location sees
the same law.  So the rate of expansion of the PWN with respect to the
pulsar is unchanged.  The question of a pulsar kick is an interesting
one, though unrelated to our present concerns with the expansion and
flux changes, but our data are not sufficient to determine the pulsar
motion at this time.

The new smaller braking index of 2.19 produces a spindown age of 1230
years and a larger value of $p$, 2.68, giving a more rapid dropoff of $L$
with time, but also a larger $\tau = t_{\rm sd} - t$.  Again
estimating with $m = 1.2$ for true ages of 430 -- 530 years, we find $L =
(0.2 - 0.3)L_0$, not inconsistent with the results for the earlier
braking index.  So the pulsar's change in $n$ does not make a large
difference in the estimated age of the system.  For a true age of 360
-- 530 years and $n$ of 2.19 or 2.65, the range of initial periods $P_0 =
P \left(L/L_0\right)^{1/{n+1}}$ is 200 -- 230 ms, and the initial
luminosity in the range $(2.5 - 7.7)L = (2 - 6) \times 10^{37}$ erg
s$^{-1}$.  These conclusions are in line with previous studies
\citep{bucciantini11,gelfand14}.

Our expansion rate corresponds to a current velocity of about 1000 km
s$^{-1}$, for a distance of 5.8 kpc.  A spherical bubble inflated
by a constant-luminosity pulsar
inside expanding uniform ejecta has after
a time $t$ a radius of
\begin{equation}
R = \left(\frac{125}{99} \frac{v_1^3 L_0}{M_c}\right)^{1/5}\, t^{6/5}
\end{equation}
where the total ejecta mass is $M_c$ with outermost expansion velocity
$v_1$ (\citealt{chevalier77}; \citetalias{rc84}). This can be rewritten in terms of
the upstream ejecta density at time $t$, $\rho_{\rm ej}(t) \propto t^{-3}$:
\begin{equation}
R = 0.79 L_0^{1/5} \rho_{\rm ej}^{-1/5} t^{3/5}.
\end{equation}
For $t = 480$ years, $R = 0.42$ pc, and assuming an intermediate value for the initial
luminosity, $L_0 \sim 4 \times 10^{37}$ erg s$^{-1}$, we obtain a
current upstream density $\rho_{\rm ej} \sim 10^{-23}$ g cm$^{-3}$,
and a swept-up mass of about 0.05 $M_\odot$.  That is, very little of
the ejecta mass has been swept up.  This rough estimate is consistent
with the determination of a low total ejecta mass ($\sim 3\ M_\odot$)
of \cite{gelfand14}, based on a similar model.  However, in contrast
to that work, we do not require any assumptions about the pulsar
behavior other than that the luminosity has not changed radically
since birth.

However, the assumption of uniform ejecta is clearly a gross
oversimplification.  While our result of a relatively low density
currently being encountered by the PWN is robust, based only on the
expansion velocity we measure and the pulsar's luminosity, we cannot
reliably infer a low total ejecta mass.  In particular, the nickel
bubble effect \citep{li93} in which energy input from radioactive
decay of $^{56}$Ni in the inner ejecta heats those ejecta, inflating a
low-density bubble, could reduce the density of the inner ejecta.
\cite{chevalier05} considers this effect in the context of PWNe, and
finds that for an initial red supergiant progenitor ejecta profile, the
density contrast between the bubble material and ambient gas is given
by
\begin{equation}
\frac{\rho_{\rm bub}\,t^3}{\rho_{\rm amb}\, t^3} = 0.052 \left( \frac{M_{\rm Ni}}{0.1 \ M_\odot}\right)^{2/5} \left( \frac{\rho_{\rm amb} t^3}{10^9 \ {\rm g\ s}^3 \ {\rm cm}^{-3}} \right)^{-2/5}.
\end{equation}
That is, ejecta beyond the bubble could have more than an order of
magnitude higher density and a much larger total mass.  So a fairly
ordinary supernova of type IIP, for which this is an appropriate
normalization density, could produce a low-density bubble in which the
PWN expands rapidly as observed.


In modeling the optical/IR spectrum of the Type IIP supernova SN
2004et, \cite{jerkstrand12} used a spherical hydrodynamic stellar
model from \cite{woosley07} but added artificial mixing in the core.
The total density in their model was $\rho\,t^3 = 1.1 \times 10^9$ g
cm$^{-3}$ s$^3$, consistent with the ambient density in Chevalier's
picture.  However, they deduced the presence of a nickel bubble with a
filling factor of about 0.15 with mean density about $9 \times
10^{-16} t_{\rm yr}^{-3}$ g cm~$^{-3}$, which would give about $7 \times
10^{-24}$ g cm$^{-3}$ at an age of 480 years.  They found that such a
model satisfactorily describes the UVOIR spectra of SN 2004et between
140 and 700 days.  The close agreement between the density of their
nickel bubble and our inferred density from the simple PWN model is
fortuitous, but the consistency of the two estimates supports the
general idea of PWN expansion into a low-density nickel bubble in the
interior of ejecta from a typical SN IIP event.

On the other hand, the idea of the origin of Kes 75 and its pulsar in
a low-energy, low ejecta mass explosion may be problematic.  Such
explosions were discussed in \cite{jerkstrand15} in a study of Type
IIb supernovae, using models with total ejecta masses below 3.5
$M_\odot$, and core masses between 0.6 and 2.3 $M_\odot$.  Good
descriptions of SN 1993J, SN 2008ax, and SN 2011dh were obtained from
the lower end of this range, with mean core densities of $7 \times
10^6$ g cm$^{-3}$ s$^3$ and $1 \times 10^7$ g cm$^{-3}$ s$^3$ for core
masses of 0.64 and 0.95 $M_\odot$, respectively.  However, only about
0.1 $M_\odot$ of the ejecta mass was made up of $^{56}$Ni, with a
filling factor of 0.6 or larger, that is, a nickel bubble with a
density lower by about an order of magnitude.  At an age of 480 years,
such a bubble would now have a density of a few times $10^{-25}$ g
cm$^{-3}$, too low for our inferred density.  While these models
certainly do not exhaust the possibilities for low-mass core
explosions, they suggest that such explanations for the Kes 75 event
may be less likely than that of a normal SN IIP event.

It is instructive to compare the Kes 75 PWN with that of SNR B0540$-$693,
inferred to have resulted from a Type IIP event.  The exhaustive study
of that object by \cite{williams08} showed that a similar model of
accelerated PWN expansion in an iron-nickel bubble could explain
observations at radio, IR, optical, and X-ray wavelengths.  SNR B0540$-$693 is
about twice the age of Kes 75; in the picture of \cite{williams08} the
age is 1140 years, and the PWN has expanded through the bubble and into
the denser shell the bubble swept up in the first few years after the
supernova.  That dense shell has fragmented into mainly oxygen-rich
clumps with a density contrast of about 100, with the PWN driving
slow, radiative shocks into them.  A simple spherically symmetric
picture analogous to that of \citetalias{rc84} or \cite{chevalier05}
gave a current
PWN shock velocity (i.e., excess of PWN bubble velocity over that of
freely expanding ejecta) of about 150 km s$^{-1}$, and a mean density
of $9 \times 10^{-24}$ g cm$^{-3}$ and total swept-up mass of about 1
$M_\odot$.  For Kes 75, we have a much smaller swept-up mass since the
current nebular radius is about one-third that in SNR B0540$-$693, but a
qualitatively similar picture can describe our observations.  Though
the extinction to Kes 75 is much too high for optical spectroscopy, IR
emission from neutral oxygen at 63 $\mu$m has been reported from
{\sl Herschel} observations\footnote{Temim, T. 2016, in Supernova Remnants: 
An Odyssey in Space after Stellar Death, id.~50, http://snr2016.astro.noa.gr.}, 
which we would attribute to
shocked clumps as in SNR B0540$-$693.  Clumping was invoked by
\cite{jerkstrand12} and would be expected for the ejecta of a Type IIP
supernova.

Given the asymmetry expected in SNe IIP events, and seen in young SNRs
such as Cas A in which iron is highly asymmetrically distributed
\citep[e.g.,][]{grefenstette17}, it is quite likely that a nickel
bubble will not be centered on the expansion center.  Such an
off-center bubble makes an attractive explanation for the asymmetry in
the PWN of Kes 75.  It could produce considerably different ambient
densities into which the two jets of the PWN are expanding (note that
our expansion measurement would apply to the western edge of the
northern half), as well as varying ejecta velocities and hence PWN
shocks.

\subsection{Flux Changes}

Our finding of a flux increase of $(5 \pm 2)$\% in the total PWN
between 2000 and 2006 is marginally consistent with either the 3\%
reported by \cite{ng08} (based only on count rates) or the
$(11^{+3}_{-4})$\% value inferred by \cite{kumar08}.  Even though our
and their values are based on absorbed fluxes, the column density is
so large ($N_{\rm H} \sim 4 \times 10^{22}$ cm$^{-2}$) that changes in
it can modify fitted fluxes by a few percent.  Our choice of \citet{grevesse98}
abundances gives very different fitted values for
$N_H$ (of order $4.6 \times 10^{22}$ cm$^{-2}$ instead of $4.0 \times
10^{22}$ cm$^{-2}$ as assumed by \citeauthor{ng08}, so differences in fitted
flux values of one or two percent are not unexpected between our
results and those of \cite{kumar08}.  However, comparison of our
observed fluxes between epochs will not suffer from this cause.


Removing the 5\% contribution to the 2006 PWN flux from scattered
pulsar X-rays would reduce our inferred PWN flux to the same within
errors as the flux from 2000, and supports our conclusion that the PWN
itself has not brightened significantly in 2006.  In addition, the
2009 flux is consistent with that from 2000.  But the drop from 2009
to 2016 of $(9 \pm 2)$\% is highly significant, and unprecedented in
X-ray studies of pulsar-wind nebulae. We stress again that these
statistical errors are always smaller than the systematic calibration
errors of 3\%, but that our results remain highly significant.

The flux decrease seems to occur over essentially the entire northern
half of the PWN.  From 2009 to 2016, the decrease in that region
(PWN-N in Figure~\ref{fig-regions}) is $(12 \pm 2)$\% or about 1.8\%
yr$^{-1}$.  A spherical nebula, powered by a more-or-less isotropic
injection of pulsar energy at a wind termination shock, can brighten
on a timescale of the sound-crossing time of the nebula $\sqrt{3}R/c$
(assuming that the nebula is dominated by relativistic fluid with
$\gamma = 4/3$ and sound speed $c/\sqrt{3}$), if the pulsar energy
input somehow increases abruptly.  However, the timescale for fading
in this one-zone picture is simply the timescale for energy loss by
adiabatic expansion or radiation; even if the pulsar suddenly ceased
its energy injection, the bubble would fade only on a dynamical
timescale (comparable to its age) or a synchrotron-loss timescale.
Abrupt lowering of the magnetic field, due to causes unknown, could
reduce the synchrotron flux, but would require the magnetic energy to
be dissipated somehow without producing any additional radiation or
other observable effect.

A more quantitative estimate of the evolutionary timescale is
possible, including the continuing pulsar energy input.  The age of
the Kes 75 system $t$ is about the same as the 
initial spindown timescale $\tau$.  Since the PWN is still encountering
unshocked ejecta, the results of
\citetalias{rc84} apply.  There it is shown that for the part of the
spectrum subject to synchrotron losses (true in the X-ray unless the
magnetic-field strength is extremely weak), the spectral luminosity
$L_x$ decreases as $L_x \propto t^l$ with the index $l$ related to the
rate of magnetic-field decrease in the bubble, $B \propto t^{-b}$, to
the injected spectral index of the initial particle distribution $s$
($N(E) = KE^{-s}$), and to the pulsar slowdown index $p$ defined
above.  The relation is
\begin{equation}
L_x \propto t^l \ \ {\rm where} \ \ l = b\left({2 - s \over 2}\right) - p.
\end{equation}
The X-ray photon index of the PWN near the pulsar is about $\Gamma
\sim 1.6$, or an energy index $\alpha_x \sim 0.6$ \citep{ng08}. We
expect a steepening due to losses of
\citep[approximately; see][]{reynolds09}
0.5 implying an injection (radio) value $\alpha_r
\sim 0.1$ or $s = 2\alpha + 1 \sim 1.2$.  \citep[This is consistent
  with the somewhat uncertain radio spectrum;][]{bock05}.  If the
magnetic energy evolved only by pulsar input and adiabatic expansion,
$b = 1.3$ at this evolutionary stage \citepalias{rc84}; magnetic
dissipation by reconnection or wave damping would cause $B$ to decline
faster, i.e., $b \ge 1.3$.  For $b = 1.3$, $s = 1.2$, and using the
pre-flare value $p = 2.21$, we find $l = -1.69$.  This rate of decline
implies a drop of 4\% in 10 years, too slow to explain our observation.
Using the post-flare value $p = 2.68$ gives $l = -2.16$ or a decline
of 5\%.  While this estimate is rough, it does indicate that the
gradual evolutionary changes expected for a young PWN cannot account
for our observations.  In any case, the absence of significant decline
between 2000 and 2006 argues strongly against any such gradual
explanation.

Can radiative losses be responsible for the flux decrease? 
The synchrotron loss timescale ($t_{1/2}$, the time for an electron
primarily radiating at frequency $\nu$ to lose half its energy) is
given by
\begin{equation}
t_{1/2} = 1.2 \times 10^3 \left(h \nu \over {1 \ {\rm keV}}\right)^{-1/2}
\left(B \over {10 \ \mu{\rm G}}\right)^{-3/2} \ {\rm years}.
\end{equation}
\cite{terrier08} estimate a magnetic-field strength of about 15
$\mu$G, based on a simple scaling of the TeV to X-ray luminosity and
the assumption that the TeV gamma-rays result from inverse-Compton
upscattering cosmic microwave background photons by relativistic
electrons (ICCMB).  However, the TeV spectrum is considerably softer
than that in X-rays \citep[$\Gamma \sim 2.3$;][]{mcbride08},
requiring intrinsic structure (i.e., not due to radiative losses) in
the electron spectrum, as well as other assumptions.  Such a low
magnetic field in an object this young is unexpected; while there is
no obvious physical reason to expect equipartition to be maintained
between particles and magnetic field, the equipartition field for the
Kes 75 PWN (based on the X-ray flux alone) is about 40 $\mu$G
\citep{ng08}, while the field strength required to produce a bend in
an originally unbroken particle spectrum at a frequency of order
$10^{15}$ Hz \citep{morton07} would be of order 100 $\mu$G.  For $B =
15\ \mu$G, the loss time for electrons predominantly radiating at 5
keV is $t_{1/2} \sim 290$ years.  Demanding $t_{1/2} \sim 10$ years implies $B
\sim 140\ \mu$G.  Since the decline we observe began only in 2006 at
the earliest, this value of $B$ would have had to be reached fairly
quickly, throughout the northern (but not southern) half of the
nebula.  This seems unreasonable.  For no decline to be observed prior
to 2006, we require $t_{1/2}$ before then to be greater than 6 years or
$B \lapprox 100\ \mu$G.  If this value characterizes the northern half
of the PWN, for a mean radius of $\sim 15''$ or $1.3 \times
10^{18}d_{5.8}$ cm, the magnetic energy content would be $U_B \sim 2
\times 10^{45}$ erg.  The maximum pulsar input in 500 years, $L_0 t \sim
(3 - 8)L(t)\, t$, is about $(0.4 - 1)  \times 10^{48}$ erg --
certainly ample to provide this pre-flare field. However, then
increasing $B$ to 140 $\mu$G in only 10 years would require an
additional $\Delta U_B \sim 2 \times 10^{45}$ erg, while at the
current rate (unchanged by the flaring) the pulsar in 10 years injects
only about this much energy.  Even if these problems were overcome,
the constancy of the photon index is not what would be expected in the
event of suddenly increased radiative losses.  Finally, such a picture
would require the radio nebula to brighten substantially.  The
synchrotron luminosity depends on magnetic field as $B^{1 + \alpha_r}$
  or about $B^{1.2}$, so a flux increase by factor of at least
  $1.4^{1.2}$ or 50\% would be expected, and could hardly be missed.

We are left with explanations relying on the inhomogeneity and
anisotropy of the PWN -- that is, on large departures from one-zone
models.  To reach the full extent of the northern nebula in 10 years
requires a signal speed $v \gapprox 0.14 c$, considerably greater than
the $0.03\,c$ inferred by \cite{ng08} for a small feature in the
southern jet.  Furthermore, the signal would need to be a decrease
rather than increase in energy input, even though substantial excess
energy was released in the 2006 flares.  More seriously, no change in
spindown luminosity of the pulsar seems to have occurred
\citep{archibald15}.

The distinct character of the northern and southern parts of the PWN
has been noted before.  Though they could not resolve substructures
such as knots and jets, \cite{bock05} pointed out that the two parts
had different properties at mm wavelengths, with the northern part
having a flatter spectrum (fainter than the south at 1.4 GHz but
brighter at 86 GHz).  They suggest additional spectral structure in
the north as well.  In the jet-torus model of \cite{ng08}, the
southern jet is approaching, making an angle of $28^\circ$ with the
plane of the sky.  However, the brightest part of the jet is the
northern knot, on the receding side.  \cite{bock05} attribute the
southern emission to shell emission seen in projection, though the
X-ray images do not seem consistent with this interpretation.
As mentioned above, an iron-nickel bubble displaced from the pulsar
could explain this asymmetry. 


\section{Conclusions}

We have detected expansion in the PWN of the composite SNR Kes 75.
Our measured PWN expansion of $(2.49 \pm 0.23)$\% in 10 years gives
free expansion ages of $400 \pm 40$ years.  Theoretical expectations
that the pulsar luminosity is still close to its birth value, and that
the PWN is encountering roughly uniform ejecta, imply acceleration of
the PWN with $R \propto t^{6/5}$, for which the true system age is
larger by 1.2, or 480 years.  We confirm directly, without recourse to
inferences based on pulsar spindown models, that Kes 75 contains the
youngest known pulsar-wind nebula in the Galaxy.


Our expansion rate implies a current expansion velocity of about
1000 km s$^{-1}$ for the PWN.  This relatively high velocity requires
a rather low density for the material into which the PWN is expanding.
While a low-energy supernova with small ejected mass cannot be ruled out,
an attractive possibility is that the PWN is expanding into a low-density
Fe-Ni bubble.  If these elements were ejected anisotropically, the
resulting asymmetric bubble could explain some of the symmetry between
the two halves of the PWN.

As has been shown for the combination remnant B0540-693 in the LMC
\citep{williams08} and for the PWN G54.1+0.3 \citep{temim10,
  gelfand15}, the PWN in Kes 75 can probe inner supernova ejecta
unobservable by other means, providing evidence on the nature of the
progenitor system.  Further observations of Kes 75, especially at
infrared wavelengths, may allow firmer conclusions to be drawn, but
what is currently known about Kes 75 seems consistent with an origin
in a fairly typical Type IIP supernova.  This finding, if confirmed,
would add to the evidence that high magnetic-field neutron stars do
not require unusual supernovae \citep[e.g.,][]{borkowski17}.

The integrated 1 -- 8 keV flux of the PWN has changed markedly since
2000.  While an apparent rise in 2006 is likely due to scattered
X-rays from the much brighter pulsar during its flaring seven days
prior to the observations, we find a decline of over 10\% in the total
PWN flux between 2000 and 2016, mainly concentrated in the northern
half of the PWN.  A bright knot there has decreased in flux by ($30
\pm 4$)\% since 2000.  These changes are well in excess of typical
calibration errors of order 3\% and smaller statistical errors and
are certainly real.

No good model exists for the fading.  One-zone models do not easily
accommodate such rapid changes, and are clearly oversimplified given
the complex structure and inhomogeneity of the Kes 75 PWN.  Properties
requiring explanation include anisotropies in energy injection into
the PWN (so the pulsar wind cannot be bilaterally symmetric), rapid
changes in brightness occurring over a large volume, and absence of a
significant change in the pulsar spindown luminosity to accompany the
fading.  The sudden change in braking index of PSR J1846$-$0258 in
2006 already pointed to difficulties with the simplest spindown
inferences \citep{archibald15}, and our detection of major PWN changes
without large changes to the pulsar spindown luminosity add to those
difficulties.  Kes 75 should continue to be monitored at X-ray and
radio wavelengths, as it may contain clues demanding significant
modifications to our ideas about pulsar winds and energy loss.

\acknowledgments We acknowledge support from NASA through {\sl
  Chandra} General Observer Program grant SAO GO6-17071X.  PHG
acknowledges support from NC State University's Provost's Professional
Experience Program.  We thank P.~Plucinsky for discussions on
systematic errors in flux calibration.  The scientific results
reported here are based on observations made by the {\sl Chandra}
X-ray Observatory.  This research has made use of software provided by
the {\sl Chandra} X-ray Center (CXC) in the applications packages {\sl
  CIAO} and {\sl ChIPS}. We acknowledge use of various open-source
software packages for Python, including Numpy, Scipy, Matplotlib,
Astropy (a community-developed core Python package for Astronomy), and
APLpy.\footnote{APLpy is an open-source plotting package for Python
  hosted at http://aplpy.github.com.}

\vspace{5mm}
\facilities{CXO}

\software{CIAO (v 4.9) \citep{fruscione06}, XSPEC \citep{arnaud96}, 
  Astropy \citep{astropy13}, matplotlib \citep{hunter07},
  Numpy \citep{walt11}, Scipy\footnote{Jones, E., Oliphant, T., Peterson, P., et al.~2001, http://www.scipy.org.}, APLpy \citep{robitaille12} }

\bibliographystyle{aasjournal}
\bibliography{pwn2}
\end{document}